\documentclass[11pt]{article}
\pdfoutput=1
\usepackage[ascii]{inputenc}
\usepackage{hyperref,dcolumn,bbm,url,booktabs,braket,microtype,aliascnt,mathtools,mathrsfs,etoolbox,cancel,capt-of,algpseudocode,float,newfloat,amsmath,amssymb,color,amsthm,fullpage,cleveref}
\usepackage[T1]{fontenc}
\usepackage[USenglish]{babel}
\usepackage{marvosym}
\usepackage{tabularx}
\usepackage{array, ltablex, multirow}

\usepackage{physics}

\usepackage{subcaption}
\usepackage{authblk}

\crefname{thm}{theorem}{theorems}
\crefname{lem}{lemma}{lemmas}
\crefname{cor}{corollary}{corollaries}
\crefname{prb}{problem}{problems}
\crefname{dfn}{definition}{definitions}

\newcommand{\ba}{\begin{eqnarray}}
\newcommand{\ea}{\end{eqnarray}}
\definecolor{armygreen}{rgb}{0.29, 0.33, 0.13}
\definecolor{applegreen}{rgb}{0.55, 0.71, 0.0}

\hypersetup{colorlinks=true,linkcolor=blue}

\begin{document}

\title{\textbf{Quantum chaos in the sparse SYK model}}

\author[1,3,5]{Patrick Orman}
\author[2,3]{Hrant Gharibyan}
\author[3,4]{John Preskill}
\affil[1]{Department of Physics, University of California, San Diego, CA 92093}
\affil[2]{BlueQubit Inc., Los Angeles, CA 90046}
\affil[3]{Institute for Quantum Information and Matter, Caltech, Pasadena, CA 91125}
\affil[4]{AWS Center for Quantum Computing, Pasadena, CA 91125}
\affil[5]{Department of Physics, University of Alabama at Birmingham, Birmingham, AL 35294}

\maketitle

\begin{abstract}
The Sachdev-Ye-Kitaev (SYK) model is a system of $N$ Majorana fermions with random interactions and strongly chaotic dynamics, which at low energy admits a holographically dual description as two-dimensional Jackiw-Teitelboim gravity. Hence the SYK model provides a toy model of quantum gravity that might be feasible to simulate with near-term quantum hardware. Motivated by the goal of reducing the resources needed for such a simulation, we study a sparsified version of the SYK model, in which interaction terms are deleted with probability $1{-p}$. Specifically, we compute numerically the spectral form factor (SFF, the Fourier transform of the Hamiltonian's eigenvalue pair correlation function) and the nearest-neighbor eigenvalue gap ratio $r$ (characterizing the distribution of gaps between consecutive eigenvalues). We find that when $p$ is greater than a transition value $p_1$, which scales as $1/N^3$, both the SFF and $r$ match the values attained by the full unsparsified model and with expectations from random matrix theory (RMT). But for $p<p_1$, deviations from unsparsified SYK and RMT occur, indicating a breakdown of holography in the highly sparsified regime. Below an even smaller value $p_2$, which also scales as $1/N^3$, even the spacing of consecutive eigenvalues differs from RMT values, signaling a complete breakdown of spectral rigidity. Our results cast doubt on the holographic interpretation of very highly sparsified SYK models obtained via machine learning using teleportation infidelity as a loss function. 
\end{abstract}


\section{Introduction}
The AdS/CFT correspondence \cite{AdSCFT1997}, a duality relating quantum gravity on a $(d+1)$-dimensional asymptotically-AdS space to a $d$-dimensional nongravitational theory defined on its boundary, has generated invaluable insights into quantum spacetime while also providing powerful tools for studying strongly-coupled quantum theories. The Sachdev-Ye-Kitaev (SYK) model \cite{SY,KitaevSYKtalks,MSRemarksOnSYK} is a particularly simple toy model of holography, which relates a system of $N$ strongly-coupled Majorana fermions with random interactions to two-dimensional Jackiw-Teitelboim (JT) gravity. The discovery of surprisingly simple holographic theories opens the possibility of exploring features of quantum gravity experimentally using relatively near-term quantum processors \cite{garcia2017digital,babbush2019quantum,QGITL1,QGITL2,bhattacharyya2022quantum, quantinuumexperiment,asaduzzaman2024sachdev}. Such quantum simulations might unveil previously unknown models with useful holographic duals, or suggest ways to understand emergent behavior in quantum matter by drawing on gravitational intuition. 

Despite its relative simplicity, simulating the SYK model on quantum hardware is a formidable task. As originally formulated, the Hamiltonian of the model contains all possible terms involving four of the $N$ Majorana fermions, with coefficients sampled from a Gaussian ensemble; hence there are $O(N^4)$ terms. However, a sparsified version of the model which is easier to simulate was proposed in \cite{SwingleSpSYK}. In the sparsified model, each term is deleted from the Hamiltonian with probability $1{-}p$ and retained with probability $p$. It was argued \cite{SwingleSpSYK} that the sparse SYK model remains strongly chaotic, and the duality to JT gravity is preserved, for $p$ scaling as $1/N^3$, in which case the mean number of nonzero terms in the Hamiltonian is $O(N)$, significantly reducing the computational resources needed to simulate the model either classically or quantumly. 

In our work we use numerical computations to investigate the impact of sparsification on the quantum chaos exhibited by the model, finding results corroborating expectations expressed in \cite{SwingleSpSYK}. For this purpose, we generate many samples of the sparsified SYK model for values of $N$ ranging from 18 to 30, and for a variety of values of $p$. For each sample, we diagonalize the Hamiltonian and compute two quantities which are useful for characterizing the eigenvalue spectrum, the spectral form factor (SFF), and the nearest-neighbor eigenvalue gap ratio. The SFF $g(t)$ is the Fourier transform of the eigenvalue pair correlation function $R(E_1,E_2)$. The gap ratio $r$ is the average over the spectrum of the ratio of two consecutive eigenvalue gaps. For the full SYK model ($p=1$) the behavior of both quantities is captured well by random matrix theory (RMT). Our computations indicate that both $g(t)$ and $r$ are robust to sparsification as long as $p$ is larger than a transition value $p_1=O(1/N^3)$. However, for $p<p_1$, we find that $g(t)$ deviates significantly from its behavior for $p=1$, signaling that in this regime the sparsified model is less strongly chaotic than the unsparsified model. Specifically, for $p<p_1$, the energy range in which $R(E_1,E_2)$ matches the predictions of RMT is reduced due to a softening of spectral rigidity. Insofar as maximally strong chaos may be viewed as necessary for the existence of a gravitational dual \cite{OnsetOfRMT}, this result indicates a breakdown of holography when $p<p_1$. We also find another transition point at $p_2<p_1$, such that for $p < p_2$ the gap ratio $r$ deviates from RMT predictions, an even more radical failure of spectral rigidity.

These conclusions apply to typical Hamiltonians with a specified value of the sparsity parameter $p$, but one may wonder whether there are special atypical Hamiltonians with the same effective value of $p$ that admit a holographic dual even though $p < p_1$. A machine learning (ML) approach for identifying such atypical Hamiltonians was proposed in \cite{experiment}. Those authors sought Hamiltonians that minimize a loss function corresponding to the infidelity of a quantum teleportation protocol for two entangled SYK-like models, where the protocol admits a gravitation dual interpretation as transmission of quantum information through a traversable spatial wormhole in the bulk \cite{gao2017traversable,maldacena2017diving}. By analyzing the SFF and the gap ratio for the models found in \cite{experiment}, we find significant deviations from the properties of the full SYK model, casting doubt on claims that these models, like unsparsified SYK, are dual to JT gravity. Our analysis complements observations in \cite{kobrin2023comment}, where the failure of these models to thermalize was noted. 
Perhaps a different ML method, seeking models such that $r$ and the SFF match their behavior for the full SYK model, could identify more promising highly sparsified holographic models that are well suited for experimental studies of quantum gravitational phenomena. 

The rest of this paper is organized as follows. In Sec.~\ref{sec:SYK} we review the SYK model and its sparse version. In Sec.~\ref{sec:random-matrices} we review relevant elements of random matrix theory and introduce the key metrics, the SFF and the eigenvalue gap ratio $r$, that are used in our analysis. In Sec.~\ref{sec:probing} we study the robustness of these metrics to sparsification, and in Sec.~\ref{sec:atypical}, we discuss features of the very highly sparsified models identified in \cite{experiment}. Sec.~\ref{sec:conclusions} contains concluding remarks, and some details regarding our computations are described in the appendices. 

\section{The SYK model and its sparsification}
\label{sec:SYK}

The Sachdev-Ye-Kitaev (SYK) model \cite{KitaevSYKtalks,MSRemarksOnSYK} is a model of $N$ Majorana fermions $\chi_a^\dag = \chi_a,\, a=1,\dots,N$, satisfying the anti-commutation relation $\acomm{\chi_a}{\chi_b}=\delta_{ab}$. Its Hamiltonian
\begin{equation}H=\frac{1}{4!}\sum_{abcd} j_{abcd}\chi_a\chi_b\chi_c\chi_d,\end{equation} 
contains all possible terms involving 4 of the $N$ fermion species; here
the coefficient $j_{abcd}$ is drawn from a Gaussian distribution with mean and variance
\begin{equation}\overline{j_{abcd}}=0, \quad \overline{j_{abcd}^2}=\frac{3!J^2}{N^3}.\end{equation} 
Models with interaction terms involving $q\ge 4$ of the $N$ fermions may also be considered, but we will confine our attention to the case $q=4$.

At inverse temperature $\beta$, the real-time two-point correlation function, averaged over the Gaussian ensemble of Hamiltonians, can be computed in the limit of large $N$ and strong coupling ($\beta J\gg 1$), yielding 
\begin{equation}
G_a(t) = \langle\chi_a(t)\chi_a(0)\rangle \propto \frac{1}{\Big(\beta \sinh(\frac{\pi t}{\beta})\Big)^{1/2}}.
\end{equation}
The exponential decay for large time $t$ signifies quantum chaos. In fact, the SYK model is maximally chaotic. The Lyapunov exponent $\lambda_L$ controlling in its out-of-time-order four-point correlator, defined by 
\begin{equation}
    \left\langle \acomm{\chi_a(0)}{\chi_b(t)}^2 \right\rangle \propto \frac{1}{N}e^{\lambda_L t},
\end{equation}
saturates the general bound $\lambda_L \leq \frac{2\pi}{\beta}$ \cite{maldacena2016bound}.

The ensemble-averaged partition function 
can also be computed for large $N$ and large $\beta J$, from which one infers that in this limit the model is accurately described by a 1D effective field theory whose effective action is given by a Schwarzian derivative \cite{KitaevSYKtalks,MSRemarksOnSYK,kitaev2018soft}. This same action provides the boundary description of 2D Jackiw-Teitelboim (JT) gravity \cite{jackiw1985lower,teitelboim1983gravitation}, establishing the holographic correspondence between the SYK model and JT gravity. 

Motivated by the desire to reduce the resources needed to simulate the SYK model using classical or quantum computers, we will consider a modified version of the model with fewer interaction terms proposed by Xu, Susskind, Su, and Swingle \cite{SwingleSpSYK}. In this sparse SYK model, interaction terms are removed from the Hamiltonian with probability $1-p$ and retained with probability $p$. In addition, to ensure that the energy and time scales dictated by the model are comparable to those of the unsparsified SYK model (with $p=1$), the variance of the Gaussian distribution from which the coefficients are drawn is rescaled by a factor $1/p$. For each coefficient $j_{abcd}$ we introduce a corresponding random variable $x_{abcd}$ taking values 0 and 1 where 
\begin{equation}
\Pr(x_{abcd}=1) \equiv p;
\end{equation}
the Hamiltonian of the sparse SYK model is 
\begin{equation}
H=\frac{1}{4!}\sum_{abcd} x_{abcd}j_{abcd}\chi_a\chi_b\chi_c\chi_d,
\end{equation}
where each Gaussian distributed $j_{abcd}$ has mean and variance\footnote{This choice of normalization is analogous to the typical SYK normalization, such that $\langle \text{Tr} H\rangle= 0$ and $\frac{1}{L}\langle \text{Tr} H^2 \rangle = N$. See \cite{OnsetOfRMT} for detailed discussion on the choice of normalization.}

\begin{equation}
\overline{j_{abcd}}=0, \quad \overline{j_{abcd}^2} = \frac{1}{p} \frac{3!J^2}{N^3}.
\end{equation}
Due to statistical fluctuations of $\{x_{abcd}\}$, the number of nonzero terms in the Hamiltonian may vary from sample to sample. 
One could enforce constraints requiring that the number of terms is the same for all sampled Hamiltonians, or that the interaction hypergraph defined by the nonzero terms is regular \cite{SwingleSpSYK}, but we will not do so here.
On average the number of nonzero terms in the Hamiltonian is $p\binom{N}{4}$; it is convenient to define a parameter
\begin{equation}
    k \equiv \frac{p \binom{N}{4}}{N} \approx \frac{pN^{3}}{24}
    \label{eq:k-p-relation}
\end{equation}
such that $k$ is an $N$-independent constant when the number of Hamiltonian terms scales linearly with the total system size $N$, which we may interpret as the average degree of the interaction hypergraph.\footnote{Here we define the ``average degree'' as the expected number of hyperedges divided by the number of vertices; the standard definition is the average number of hyperedges containing each vertex, which is 4 times larger.} It is suggested in \cite{SwingleSpSYK}, based partly on numerical results, that the
the minimal value of $k$ required for gravitational physics to emerge at low temperature lies between $1/4$ and $4$. We will not study directly the gravitational interpretation of the sparse SYK model; rather we will investigate instead the range in $k$ for which the model exhibits maximal chaos, which we view as a necessary condition for the existence of a gravitational dual. Our study leverages tools from random matrix theory, reviewed in Sec.~\ref{sec:random-matrices}.

\section{Quantum chaos and random matrices} 
\label{sec:random-matrices}

Reflecting the model's strongly chaotic character, the energy eigenvalue spectrum of the SYK model has features that are well described by random matrix theory (RMT) \cite{BHandRMT,saad2018semiclassical,OnsetOfRMT}, which we now briefly review. Indeed, we expect RMT to capture universal properties of the spectrum for generic chaotic quantum systems, with the appropriate matrix ensemble dictated by the symmetries of the system. To be concrete, consider the Gaussian unitary ensemble (GUE) \cite{dyson1962brownian,mehta1986random}, which has the partition function 
\begin{equation}\label{GUE-measure}
\mathcal Z_{\rm GUE} = \int \left(\prod_{i,j}dM_{ij}\right)e^{-\frac{L}{2}{\rm Tr}M^2}, 
\end{equation}
where $M$ is an $L\times L$ Hermitian matrix. We use $\{E_i\}$ to denote eigenvalues of the matrix $M$ listed in ascending order, and define the normalized eigenvalue density $\rho(E) = \frac{1}{L} \sum_i^{L} \delta(E - E_i)$. For large $L$, the average density in this ensemble is given by the Wigner semicircle law
\begin{equation}\label{rhosemicircle}
\langle \rho (E) \rangle_{\rm GUE}=  \rho_{\rm sc}(E) =\frac{1}{2\pi} \sqrt{4-E^2}. 
\end{equation}
With this normalization, the entire energy band is supported on an interval of constant width, and hence the typical gap between adjacent eigenvalues scales like $1/L$. Correlations among the eigenvalues are captured by the eigenvalue pair correlation function $R(E_1, E_2) \equiv \langle \rho(E_1) \rho(E_2)\rangle $. 
When $\bar E=\frac{1}{2}(E_1+E_2)$ is not close to the band edge and the separation $|E_1-E_2|$ is much less than 1, $R(E_1,E_2)$ is given by the sine kernel formula \cite{mehta1986random} 

\begin{equation}\label{sine-kernel}
R(E_1, E_2) 
 = \frac{ \rho_{\rm sc}(E_1)}{\pi L} \delta(E_1 -E_2) +\rho_{\rm sc}(E_1)\rho_{\rm sc}(E_2)\left(1- \frac{\sin^2 x}{x^2} \right);
\end{equation}
here for convenience we have defined $x \equiv \pi L (E_1 -E_2)\rho({\bar E})$.

When $|E_1 -E_2| \gg 1/L$ (energy difference large compared to the mean spacing between eigenvalues), the amplitude of the wiggles in the pair correlation function falls off as $\frac{1}{x^2}$. This characteristic dampening of the long-range fluctuations in the eigenvalue ``gas'' is known as ``spectral rigidity.'' Although spectral rigidity is derived in the unphysical setting of RMT, it is also observed in actual chaotic quantum Hamiltonian systems; hence we speak of \emph{universal} RMT behavior. More precisely, spectral rigidity is applicable
when the energy difference is less than a characteristic value called the Thouless energy, while spectral rigidity breaks down when $|E_1-E_2|$ is greater than the Thouless energy. The time scale inversely related to the Thouless energy is called the Thouless time. It has also been called the \emph{ramp time}, for reasons we explain below.

\subsection{Spectral form factor}

A commonly used diagnostic for spectral rigidity is the \emph{spectral form factor} (SFF) \cite{mehta1986random,guhr1998random}, which has been applied to the SYK model and other holographic theories in \cite{BHandRMT,saad2018semiclassical,OnsetOfRMT,Garcia-Garcia,xu2021thermofield,cornelius2022spectral,caceres2022spectral,matsoukas2023non}. 
The SFF (ensemble averaged at infinite temperature) is defined as
\begin{equation} \label{eq:SFF} 
g(t) = \frac{\langle |\Tr e^{-iHt}|^2\rangle}{L^2} = \frac{\langle\sum_{ij} e^{i(E_i-E_j)t}\rangle}{L^2} = \frac{1}{L^2}\int dE_1 dE_2\, L^2 R(E_1, E_2) e^{i(E_1-E_2)t};
\end{equation}
it is the Fourier transform of the pair correlation function $R(E_1,E_2)$. 

When the time $t$ is larger than the Thouless time, the energy integral in Eq.~\eqref{eq:SFF} is dominated by the regime in which the energy difference $|E_1-E_2|$ is less than the Thouless energy. Therefore we can use RMT to compute the SFF. If we make the further approximation of replacing $\rho_{\rm sc}(E)$ in Eq.~\eqref{rhosemicircle} with a constant and extend the integration ranges to $(-\infty, \infty)$, then the Fourier transform of Eq.~\eqref{sine-kernel} can be calculated, yielding 
\begin{equation}\label{rampplateau}
g(t) \sim 
\begin{cases}
t/(2 \pi), &t < 2L,\\
L/\pi,     &t \geq 2L.
\end{cases}
\end{equation}

For many-body chaotic quantum systems, the general shape of the function $g(t)$ has three distinctive regimes. When the time $t$ is less than the Thouless time, $g(t)$ drops sharply. This portion of the plot is nonuniversal; that is, it depends on features of the eigenvalue spectrum that vary from one model to another. Once $t$ reaches the Thouless time, however, the universal behavior predicted by RMT kicks in and Eq.~\eqref{rampplateau} applies. Starting here, $g(t)$ increases linearly, a regime called the ``ramp.'' When $t$ is larger than the inverse of the typical spacing between eigenvalues, the upward ramp ends, giving way to a horizontal plateau \cite{OnsetOfRMT,BHandRMT}.

To interpret these universal aspects of the SFF at sufficiently large $t$, we recall that eigenvalues of generic matrices repel. The plateau reflects the strong repulsion when energy differences are comparable to the average spacing between consecutive eigenvalues. The ramp arises because eigenvalue repulsion persists even for eigenvalue pairs that are widely separated, and as a result eigenvalue fluctuations in the GUE ensemble are much smaller than would be expected in the absence of this repulsion. Thus the ramp is a manifestation of long-range spectral rigidity. 

We refer to the time when $g(t)$ begins ramping upward as the \textit{ramp time} $t_\text{ramp}$; it is essentially the same as the Thouless time. We will investigate how $t_\text{ramp}$ depends on the sparsity parameter $p$ of the sparse SYK model, finding that the onset of the ramp is robust with respect to random deletion of interaction terms until $p$ is sufficiently small. Once $p$ falls below a transition value $p_1$, though, the ramp time exceeds its value in the full SYK model. This growth of $t_\text{ramp}$ indicates that when the interaction terms become sufficiently sparse, the spectral rigidity of the SYK model is compromised. We interpret this deviation from maximal chaos as evidence that the holographic correspondence with JT gravity is also disrupted \cite{Mertens2023}. 

We conduct our studies of spectral rigidity by computing the SFF at infinite temperature, yet we draw conclusions concerning the emergence of JT gravity at low temperature. Our attitude is that the infinite-temperature SFF for the unsparsified SYK model reveals universal features of eigenvalue repulsion that apply across the full spectrum, including at low energy. Semiclassical calculations in JT gravity that recover the linear ramp of the SFF \cite{saad2018semiclassical} (a signature of universal eigenvalue statistics) support this viewpoint. Therefore, significant deviations from RMT predictions in highly sparsified SYK models strongly suggest that these models behave much differently than the unsparsified model, including in the low-energy regime.

\subsection{Gaussian-filtered spectral form factor}
\label{subsec:filtering}

\begin{figure}[ht]
    \centering
    
    \begin{subcaptiongroup}
    \centering
    \parbox{0.48\textwidth}{\centering \includegraphics{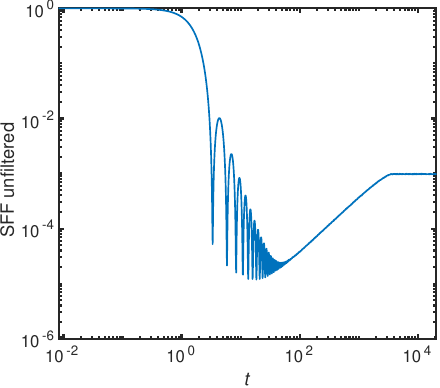} \caption{Unfiltered SFF, $\alpha=0$}}
    \parbox{0.48\textwidth}{\centering \includegraphics{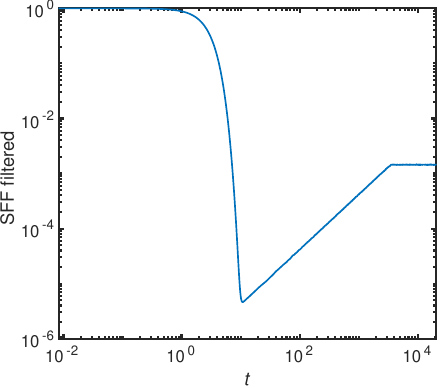} \caption{Gaussian-filtered SFF, $\alpha=3.4$}}
    \end{subcaptiongroup}
    
    \caption{Comparison of the conventional SFF and the Gaussian-filtered SFF for the full ($p=1$) SYK model with $N=22$, averaged over 20,000 samples.}
    \label{fig:SFF-filtering-comparison}
    
\end{figure}

Unfortunately, nonuniversal features of the energy spectrum give rise to large oscillations in the SFF near the ramp time, making it hard to identify a precise time when the ramp begins. We circumvent this difficulty by introducing a filter function that tames these oscillations.\footnote{This modification to the SFF was suggested by Douglas Stanford and discussed in \cite{OnsetOfRMT}. See also \cite{cornelius2022spectral}.} We multiply the energy spectrum by a Gaussian function (suppressing its sharp edges), obtaining
\begin{equation} 
\label{eq:GaussFiltSFF} |Y(\alpha,t)|^2 =  \left| \textstyle \sum_i e^{-\alpha E_i^2} e^{-iE_i t}\right|^2 ,
\end{equation}
and consider
\begin{equation}h(\alpha,t) = \left\langle \frac{|Y(\alpha,t)|^2}{|Y(\alpha,0)|^2} \right\rangle.\end{equation}
We tune the filtering parameter $\alpha$, which controls the width of the Gaussian, until the clearest/earliest ramp is seen.\footnote{See Appendix \ref{appendix:alphachoice} for discussion concerning how $\alpha$ is chosen.} 
(The optimal $\alpha$ depends only on $N$.) Setting $\alpha=0$ yields the unfiltered SFF $g(t)$. How Gaussian filtering smooths out the oscillations of the SFF for the unsparsified SYK model is illustrated in Fig.~\ref{fig:SFF-filtering-comparison}.

\subsection{Nearest-neighbor gap ratio}

The ramp exhibited by the SFF characterizes spectral rigidity for large energy differences, while the plateau arises from eigenvalue repulsion when energy differences are small, comparable to the average eigenvalue spacing $1/L$. Another useful diagnostic for spectral rigidity at such small energy differences is the universal probability distribution predicted by RMT for the interval between consecutive eigenvalues $s_i=E_i- E_{i-1}$. For GUE at large $L$, this distribution is the Wigner surmise
\begin{equation}
P(s) = \frac{32}{\pi^2} s^2 e^{-\frac{4}{\pi}s^2}.
\end{equation}
One can test for quantum chaos by checking whether $P(s)$ applies to the observed eigenvalue spacing, but this test can be tricky to apply due to fluctuations in the local eigenvalue density. 
We will instead test for chaos using a different metric (proposed in \cite{oganesyan2007localization}) 
which is more accessible from numerical computations of the spectrum. This is the \textit{nearest-neighbor gap ratio} $r$,  defined as the average over the spectrum of the ratios of two successive eigenvalue gaps:
\begin{equation}r = \left\langle \min \left(\frac{s_i}{s_{i+1}}, \frac{s_{i+1}}{s_i} \right) \right\rangle_{i}= \left\langle \min \left(\frac{E_i - E_{i-1}}{E_{i+1} - E_{i}}, \frac{E_{i+1} - E_{i}}{E_i - E_{i-1}} \right) \right\rangle_{i}, \end{equation}
where the lowest ratio is always taken (\emph{i.e.}, the ratios are always at most 1). If there were no eigenvalue repulsion, and spacings were sampled independently from the Poisson distribution, one could compute that the gap ratio $x$ samples from the distribution $Q_\text{Poisson}(x)= 2/(1+x)^2$, which when averaged yields the value $r=2\ln 2-1\approx .386$. Repulsion pushes the value of $r$ upward compared to this Poisson value.

As explained in \cite{you2017sachdev}, the RMT predictions for the spectrum of the SYK model depend on the value of $N$ modulo 8, where $N$ (an even integer) is the number of Majorana fermion species and $L=2^{N/2}$ is the Hilbert space dimension. This happens because the model has a particle-hole symmetry, and how this symmetry acts on the even and odd parity sectors depends on $N$. The upshot is that the appropriate matrix ensemble is GUE when $N ~\textrm{mod} ~8$ is 2 or 6, GOE when $N ~\textrm{mod} ~8$ is 0, and GSE when $N ~\textrm{mod} ~8$ is 4. The gap ratio $r$ has been computed for each of these matrix ensembles, both analytically in the large-$L$ limit and numerically for various values of $L$ in \cite{rLitVals,rBlocksVals}. As reported in Table~\ref{tab:r-litvals}, we find good agreement between these RMT predictions and our results for the full ($p=1$) SYK model for values of $N$ that we investigated. To obtain this agreement, it is important to block diagonalize the Hamiltonian and analyze the eigenvalue gaps in a particular symmetry sector. (Our values of $r$ were obtained by averaging over samples in the SYK model, but we note that $r$ actually varies little from sample to sample.) We wish to investigate whether this agreement with expectations from RMT survives when the SYK model is sparsified. 

\begin{table}[ht]
    \centering
    \begin{tabular}{|c|c|c|c|}
    \hline
    $N ~\textrm{mod} ~8$ & 2, 6 & 4 & 0 \\
    \hline
    Ensemble & GUE & GSE & GOE \\
    \hline
    $r$ (analytic) & 0.60266 & 0.67617 & 0.53590 \\
    \hline
     $r$ (numerical) & 0.5996(1) & 0.6744(1) & 0.5307(1)\\
    \hline
   $r$, 2 blocks (analytic) & 0.422085 & 0.411762 &  0.423415\\
    \hline
    $r$, 2 blocks (numerical) & 0.4220(5) & 0.4116(5) & 0.4235(5)\\
    \hline
    $r$, SYK model (numerical, $p=1$) & 0.599700 & 0.411057 & 0.423406 \\
    \hline
    $r$, SYK model (numerical, $p=.01$) & 0.599733 & 0.410879 & 0.423598 \\
    \hline
    \end{tabular}
    
    \caption{Values of the gap ratio $r$, for the three random matrix ensembles, obtained analytically and numerically in \cite{rLitVals,rBlocksVals}. Our numerically computed values for the ($p=1$) SYK model align with these random-matrix-theory predictions; for the GSE and GOE ensembles, we compare with the case where the Hamiltonian decomposes into two symmetry sectors, corresponding to the parity sectors in the SYK model. We list the gap ratios obtained in our numerics for the highest $N$ value computed for each ensemble ($N=30,\,28,\,24$), and for two values of $p$, averaged over all samples. Computed values of $r$ for the sparse model with $p=.01$ agree with those obtained for the unsparsified model.}
    \label{tab:r-litvals}
    
\end{table}

\begin{figure}[b!]
    \centering
    
    \begin{subcaptiongroup}
    \centering
    \parbox{0.48\textwidth}{\centering \includegraphics{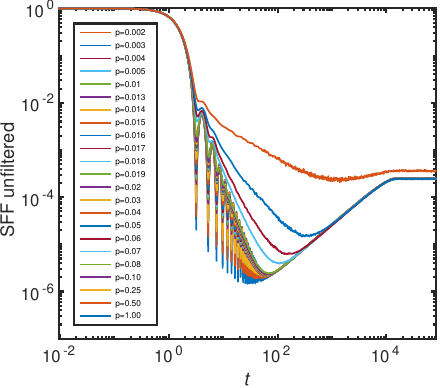} \caption{Unfiltered SFF, $\alpha=0$}}
    \parbox{0.48\textwidth}{\centering \includegraphics{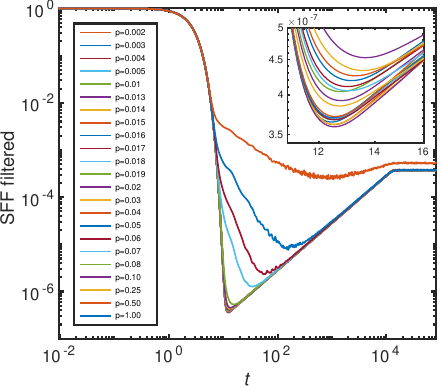} \caption{Gaussian-filtered SFF, $\alpha=3.2$}}
    \end{subcaptiongroup}
    
    \caption{Spectral form factor (a) without and (b) with Gaussian filtering, averaged over 12,000 samples, for $N=26$ and various values of $p$.}
    \label{fig:SFF-YY-N26}
    
\end{figure}

\section{Probing quantum chaos in the sparse SYK model} 
\label{sec:probing}

First we consider the SFF of the sparse SYK model for $N=26$, where $L=2^{13}=8192$ and the appropriate RMT ensemble is GUE. The SFF is plotted as a function of $t$ (a log-log plot) in Fig.~\ref{fig:SFF-YY-N26}, with and without Gaussian filtering, for various values of $p$. In the filtered plot, we find that the SFF is stable for a while as $p$ decreases, but eventually, for sufficiently small $p$, significant deviations from the SFF of the full SYK model appear. At that point, the ramp time $t_\text{ramp}$ begins to increase as $p$ decreases further. This growth of the ramp (Thouless) time indicates that the eigenvalue difference regime in which spectral rigidity applies is shrinking. At still smaller values of $p$, the ramp disappears completely, and no hint of spectral rigidity survives.

\begin{figure}[ht]
    \centering
    
    \includegraphics{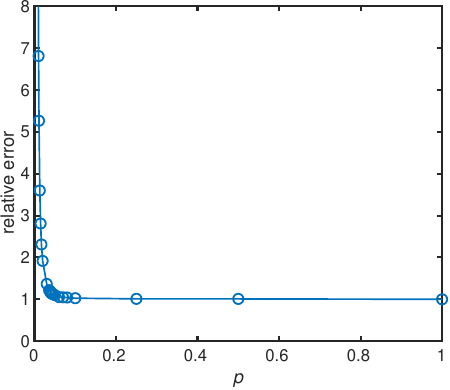}
    \caption{The relative error ($\min(YY^*_{p}) / \min(YY^*_{p=1})$) for various values of $p$ in the sparse SYK model with $N=20$, using a Gaussian filter with $\alpha=3.6$. 
    This value of the SFF at the onset of the ramp is seen to be highly resilient when the model is sparsified.}
    \label{fig:RelErrvsP-example-N20}
    
\end{figure}

To quantify the deviation of the SFF in the sparse model from its counterpart in the unsparsified model, we consider the ``relative error'' at the onset of the ramp, defined as the ratio $\min(YY^*_{p}) / \min(YY^*_{p=1})$. In Fig.~\ref{fig:RelErrvsP-example-N20} we plot this relative error as a function of $p$ for $N=20$. It is evident that according to this metric the SFF of the SYK model is quite resilient to sparsity. Even when over 90\% of the interaction terms in the Hamiltonian are deleted, the minimum value of the SFF is still nearly identical to that of the full SYK. This resilience 
was noticed earlier \cite{SwingleSpSYK,Garcia-Garcia} using other diagnostic criteria. We also note that the computational resources needed to compute the SFF are considerably reduced in sparse models. For example, in the case $N=20$ we sampled 40,000 different Hamiltonians for each value of $p$. The time needed to diagonalize all of the sampled Hamiltonians was 27 minutes for $p=0.10$ compared to 5 hours for the full SYK model with $p=1$.

\subsection{\boldmath Deformation of the ramp: the transition point $p_1$}\label{subsec:deform-graph}

\begin{figure}[ht!]
    \centering
    
    \begin{subcaptiongroup}
    \centering
    \parbox{0.48\textwidth}{\centering \includegraphics{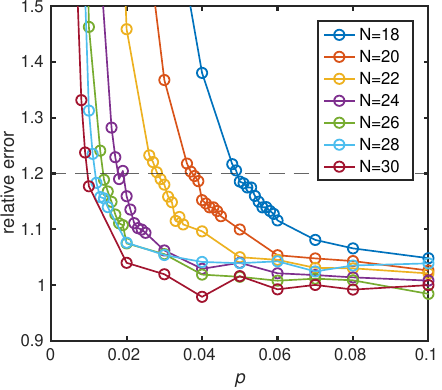} \caption{Relative error vs.\ $p$}}
    \parbox{0.48\textwidth}{\centering \includegraphics{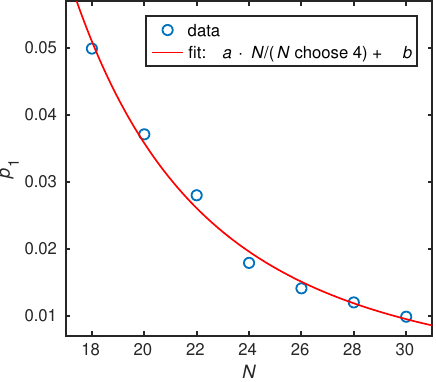} \caption{Estimated $p_1$ vs.\ $N$ and fit}}
    \end{subcaptiongroup}
    
    \caption{
    (a) Relative error vs.~$p$ for values of  $N$ from 18 to 30. 
    The horizontal dashed line shows relative error $1.2$, our threshold for determining the transition point $p_1$.
    (b) Estimated $p_1$ vs.~$N$. The red curve is the fit to $p_1=k_1N/\binom{N}{4}$ used to determine the average degree of the interaction hypergraph $k_1\approx 8.7$.
    }
    \label{fig:p1-vs-N-sidebyside}
    
\end{figure}

We denote by $p_1$ the largest value of $p$ for which the SFF of the sparse model deviates significantly from the SFF of the unsparsified model. To pinpoint this value we use the relative error at the minimum of the SFF. Somewhat arbitrarily, we define $p_1$ as the value of $p$ for which the relative error is 1.2. Fortunately, because the deviation of the sparse SFF from the unsparsified SFF kicks in very abruptly, our findings are quite insensitive to this choice. 

The relative error is depicted in Fig.~\ref{fig:p1-vs-N-sidebyside}a for values of $N$ ranging from 18 to 30, along with the $1.20$ threshold. Based on this data, we plot $p_1$ as a function of $N$ in Fig.~\ref{fig:p1-vs-N-sidebyside}b. We find a good fit to the expression
\begin{equation}
p_1 = \frac{k_1N}{\binom{N}{4}}\approx \frac{24 k_1}{N^3},
\end{equation}
where $k_1\approx 8.7$ is a constant, which we may interpret as the average degree of the interaction hypergraph for sparsity $p_1$. We note that this value of $k_1$ is larger than the upper bound conjectured in \cite{SwingleSpSYK} ($k_\text{min} < 4$) on the average degree such that, in the large-$N$ limit, Schwarzian dynamics emerges at low temperature for $k> k_\text{min}$. Our results might not be inconsistent with \cite{SwingleSpSYK}; 
conceivably, for sparse models in the region $k_\text{min} < k<k_1$, the holographic correspondence with JT gravity could still hold, even though the onset of the ramp is delayed compared to the full SYK model. 

However \cite{anegawa2023sparse} raises a further puzzle. By performing a saddle point analysis in the large-$N$ limit, those authors claimed that a Schwarzian mode described by JT gravity emerges at low energy in sparse SYK when the number of terms in the Hamiltonian is superlinear in $N$, but not when the number of terms is linear in $N$. We are not sure how to reconcile that finding with the conclusion in our work and in \cite{SwingleSpSYK} that sparse SYK behaves like unsparsified SYK when $k$ is a sufficiently large constant (and hence the number of terms is linear in $N$). It may be that $k_1$ is not really constant, but instead grows without bound as $N$ increases, where the growth is too slow to be detected in our numerical studies. 

It should be noted, though, that \cite{anegawa2023sparse} considered the average over samples of the partition function of the sparse SYK model, rather than the sample average of the free energy (the logarithm of the partition function), which is of more direct physical interest. In the unsparsified SYK model, this ``mean-field'' computation can be justified \cite{kitaev2018soft}, and it correctly yields the Schwarzian action in the large-$N$ limit. It is possible, though, that the mean-field approximation is inaccurate in the large-$N$ limit of the sparsified model with a fixed value of $k$, and therefore fails to reveal the Schwarzian mode even though it is actually present. For this reason, we do not consider the conclusions of \cite{anegawa2023sparse} to be definitive.

\subsection{\boldmath Disappearance of the ramp: the transition point $p_2$}

As $p<p_1$ continues to decrease, the SFF deviates more and more from the SFF of the unsparsified model, as illustrated in Fig.~\ref{fig:SFF-YY-N26}. Eventually, at a value of $p$ we denote as $p_2$, the ramp disappears completely, signifying the loss of spectral rigidity at all energy difference scales. From Fig.~\ref{fig:SFF-YY-N26} we may infer $p_2\approx .002$ for $N=26$.

Since we find it difficult to identify a precise value of $p_2$ from plots of the SFF, we use a different criterion instead: the eigenvalue gap ratio $r$, which characterizes eigenvalue repulsion for energy difference comparable to the average spacing between eigenvalues.  We find that the value of $r$ is robust against sparsification until $p$ is reduced to $p_2$, at which point it turns sharply downward due to the breakdown of spectral rigidity at small energy difference scales. 

\begin{figure}[!b]
    \centering
    
    \begin{subcaptiongroup}
    \centering
    \parbox{0.48\textwidth}{\centering \includegraphics{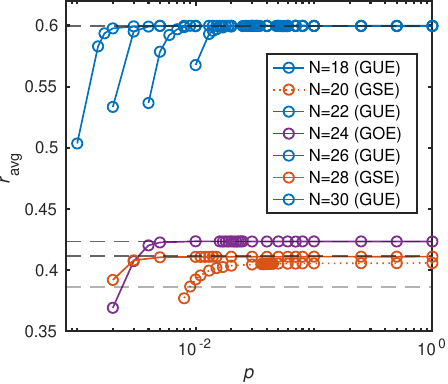} \caption{Gap ratio $r$ vs.\ $p$}}
    \parbox{0.48\textwidth}{\centering \includegraphics{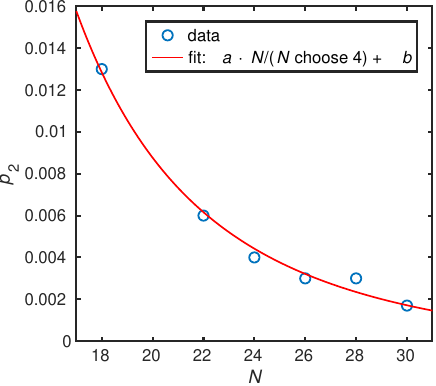} \caption{Estimated $p_2$ vs.\ $N$ and fit}}
    \end{subcaptiongroup}
    
    \caption{
    (a) Nearest-neighbor gap ratio $r$ vs.~$p$, for $N$ ranging from 18 to 30. For sufficiently large $p$, $r$ agrees well with RMT predictions (with the exception of $N=20$, the smallest value of $N$ for which the matrix ensemble is GSE). The RMT predictions are marked with dashed lines, and the Poisson value is marked with a lighter dashed line. The value of $p_2$, where $r$ drops sharply below the RMT prediction, occurs at significantly lower sparsity than $p_1$.
    (b) Value of $p_2$ vs.~$N$ and the associated scaling. 
    }
    \label{fig:p2-vs-N-sidebyside}
    
\end{figure}

We plot $r$ as a function of $p$ for $N$ ranging from 18 to 30 in Fig.~\ref{fig:p2-vs-N-sidebyside}a. For almost every value of $N$, we find that when $p$ is large enough the observed value of $r$ agrees well with RMT predictions \cite{rLitVals,rBlocksVals} for the appropriate matrix ensemble (GUE, GOE, or GSE). There is one exception --- for $N=20$ our numerics yields $r = 0.4055$, significantly below the RMT prediction $r = 0.4116$ for the large-$L$ limit of GSE. From this we infer that GSE is more susceptible to small-$L$ effects than the other ensembles, and we therefore omit $N=20$ from our analysis of how $p_2$ depends on $N$. 

We see that $r$ breaks sharply downward when $p$ is sufficiently small, and we identify $p_2$ as the value of $p$ where $r$ reaches $99\%$ of its value for $p=1$. We compute $r$ from the raw eigenvalues themselves, rather than subjecting them to block diagonalization to account for parity sectors --- leaving the eigenvalues raw leads to computed values for GSE and GOE ensembles agreeing with those in \cite{rBlocksVals} in the case of two blocks. 

We plot $p_2$ as a function of $N$ in Fig.~\ref{fig:p2-vs-N-sidebyside}b. This plot fits well with
\begin{equation}
p_2 = \frac{k_2N}{\binom{N}{4}}\approx \frac{24 k_2}{N^3},
\end{equation}
where $k_2\approx 2.3$ is a constant, which we may interpret as the average degree of the interaction hypergraph for sparsity $p_2$. Combining with our conclusions regarding $p_1$, we infer that the rigidity of the eigenvalue spectrum matches that of the unsparsified model when the average degree is $k> k_1\approx 8.7$, that it deviates notably from the unsparsified model when energy differences are large for $k_2 < k < k_1$, and that it deviates significantly even for small energy differences for $k< k_2$. 

Revisiting the comparison between our findings and those of \cite{SwingleSpSYK}, where it is suggested that $1/4 < k_{\text{min}} < 4$, the precise relationship between $k_{\text{min}}$ and $k_2 \approx 2.3$ remains uncertain. The scenario where $k_{\text{min}} < k_2$ appears unlikely, as it would suggest a regime where the system's nearest neighbor repulsion vanishes, yet the duality with JT gravity persists --- an improbable situation. On the other hand, if $k_{\text{min}} > k_2$, it implies that the JT gravity duality fails before reaching $k_2$, indicating a transition away from the standard gravitational duality before the complete disappearance of nearest-neighbor repulsion.  This issue might be resolved by refining the analysis in \cite{SwingleSpSYK} to determine $k_{\text{min}}$ more precisely, or by deriving a tighter lower bound.

\begin{figure}[b!]
    \centering
    
    \begin{subcaptiongroup}
    \centering
    \parbox{0.48\textwidth}{\centering \includegraphics{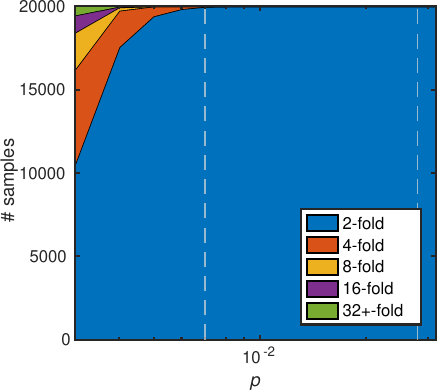} \caption{Number of samples vs.\ $p$}}
    \parbox{0.48\textwidth}{\centering \includegraphics{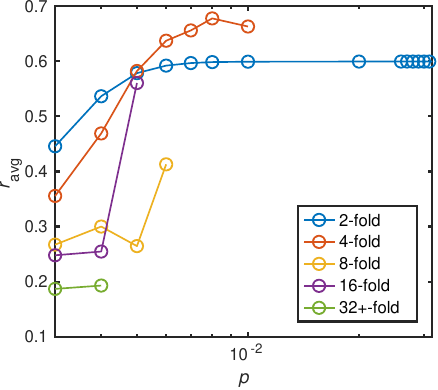} \caption{Gap ratio $r$ vs.\ $p$}}
    \end{subcaptiongroup}
    
    \caption{(a) As a function of $p$, the number of samples that have spectral degeneracies ranging from 2-fold to 32-fold and beyond for $N=22$. The transition points $p_1$ and $p_2$ are indicated by vertical dashed lines.
    (b) The ensemble-averaged value of the gap ratio $r$ for each degeneracy class. 
    }
    \label{fig:example-degenemergence}
    
\end{figure}

We also find that for $p<p_2$ many samples have surprisingly high spectral degeneracies, as shown in Fig.~\ref{fig:example-degenemergence}a. These degeneracies were also noticed in \cite{Garcia-Garcia}, and explained there in terms of emergent symmetry.
As indicated in Fig.~\ref{fig:example-degenemergence}b, each degeneracy class has a distinct value for the ensemble-averaged gap ratio $r$. However, for the data plotted in Fig.~\ref{fig:p2-vs-N-sidebyside} we included only the samples whose degeneracy matched that of the unsparsified model in the same symmetry class. 

We note that for a $d$-uniform random hypergraph, with $N$ vertices and $kN$ hyperedges where $k> 1/(d(d-1))$, the hypergraph has a unique ``giant'' connected component containing $\Theta(N)$ vertices with high probability \cite{schmidt1985component,karonski2002phase}. Therefore the transitions at $k_1,k_2$ occur in the regime ($k>1/12)$, where the interaction hypergraph is very likely to contain a giant connected component. The transition in spectral rigidity seems to be unrelated to the phase transition in the hypergraph topology.

\subsection{Dependence of ramp time on sparsity}

Instead of regarding the relative error at the minimum of the SFF as a proxy for the ramp time $t_\text{ramp}$, as we did in Sec.~\ref{subsec:deform-graph}, we can 
estimate how $t_\text{ramp}$ itself varies as a function of $p$ using other criteria. Numerically, this is most safely analyzed for the ensemble that has a completely linear ramp (namely GUE \cite{BHandRMT}), which in our case would be for $N = 18,\, 22,\, 26,\, 30$. To define the ramp time in a sparse SYK model, we might choose the time $t$ at which the SFF of the sparse model intersects with the linear ramp of the unsparsified model \cite{OnsetOfRMT}. However, because large sample-to-sample fluctuations in the SFF cause problems for this approach, we instead use a threshold relative error to define $t_\text{ramp}$.

\begin{figure}[b!]
    \centering

    \begin{subcaptiongroup}
    \centering
    \parbox{0.48\textwidth}{\centering \includegraphics{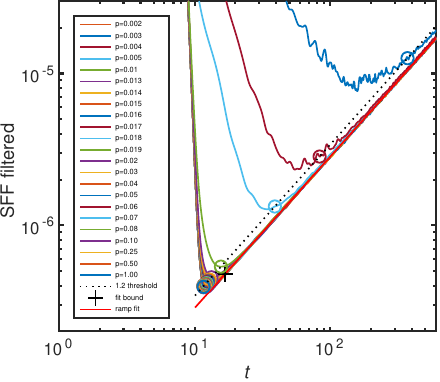} \caption{SFF vs.~$t$, and $t_\text{ramp}$ choice}}
    \parbox{0.48\textwidth}{\centering \includegraphics{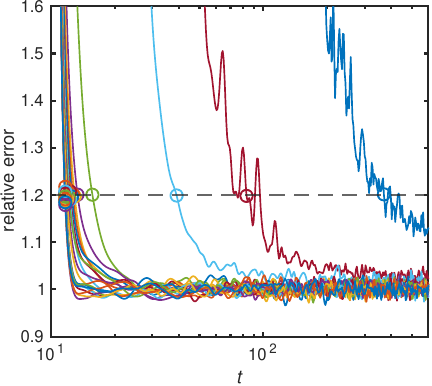} \caption{Relative error vs.~$t$, and $t_\text{ramp}$ choice}}
    \end{subcaptiongroup}

    \caption{Our method for identifying $t_\text{ramp}$ in the spectral form factor for $N=26$ and various $p$ values.
    In (a), the 20\% relative error threshold is illustrated as a dotted line parallel to the $p=1$ ramp, and each $p$ value's choice of $t_\text{ramp}$ is indicated by a circle. 
    In (b), we plot the ratio of the SFF to its estimated value determined by a linear fit to the ramp of the $p=1$ model, with the relative error threshold indicated by the horizontal line at ratio 1.2. 
    }
    \label{fig:t_ramp-choice} 
    
\end{figure}

\begin{figure}[ht]
    \centering
    
    \includegraphics{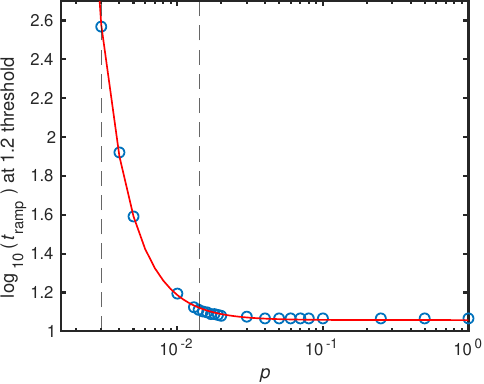}
    
    \caption{Ramp time $t_\text{ramp}$ vs.~$p$ for $N=26$. For $p$ in between $p_1$ and $p_2$ (indicated by vertical lines), the plot fits a curve of the form $t_\text{ramp} \approx a/p^c + b$, where $a\approx 1.02 \times 10^{-5}$, $b\approx1.06$, and  $c \approx2.05$.}
    \label{fig:t_ramp-vs-p-N26}
    
\end{figure}

We use a numerical approach originally proposed in \cite{OnsetOfRMT} to estimate the ramp time. We first apply a linear fit to the ramp of the SFF for unsparsified SYK, and then choose $t_\text{ramp}$ to be the value of $t$ at which the SFF for the sparse model is 20\% greater than this line. An illustration of this procedure is shown for $N=26$ in Figs.~\ref{fig:t_ramp-choice}a (SFF) and \ref{fig:t_ramp-choice}b (relative error). We find that this threshold is large enough so that the onset of the ramp is not obscured by statistical fluctuations, which is consistent with the results of \cite{OnsetOfRMT} for the full SYK model.

We plot $t_\text{ramp}$ versus $p$ in Fig.~\ref{fig:t_ramp-vs-p-N26} for $N=26$. We observe that  $t_\text{ramp}$ is nearly independent of $p$ for $p>p_1$, and fits a curve of the form $t_\text{ramp}\approx a/p^2 + b$ for $p$ between $p_2$ and $p_1$. 
We are not sure how to explain this $1/p^2$ scaling of the ramp time, but we suspect it can be derived by analyzing diffusion of energy on the interaction graph of the sparse SYK model. A similar phenomenon was discussed in \cite{OnsetOfRMT} for models with conserved charges, and we anticipate that a related hydrodynamic picture applies to energy diffusion in the SYK model. Further, an intriguing direction for future research involves exploring a sparse variant of the XXZ circuit as described in \cite{OnsetOfRMT}, with the goal of developing an analytical framework that explains the $1/p^2$ scaling of ramp time. 
This exploration might illuminate how the ramp time (and onset of universal chaos) is related to emergent hydrodynamics.

\section{Quantum chaos for atypical sparsified SYK models}
\label{sec:atypical}

Our analysis of chaos in Sec.~\ref{sec:probing} applies to typical sparse SYK models with sparsity parameter $p$. Our main conclusion is that the spectral rigidity of the model is softened for $p<p_1$ (and correspondingly for average degree $k < k_1$), which we interpret as evidence that the holographic correspondence with JT gravity is also disrupted in this regime. The spectral rigidity is completely demolished for $p<p_2$ (and $k< k_2)$. For those who wish to investigate properties of JT gravity by simulating its dual SYK model on classical or quantum computers, the lesson is that we can reduce the computational resources substantially by sparsifying the SYK model, but we should not go so far as to choose $p < p_1$, and even more so should avoid the regime with $p< p_2$.

However, this conclusion applies to the \textit{typical} models with sparsity $p$, leaving open the possibility that an even less costly simulation could be conducted by studying just the right sparse model (or models) that accurately correspond(s) to JT gravity even though the effective level of sparsity is less than $p_1$. This goal motivated the authors of \cite{experiment} to propose extremely sparse models with $N=10$ Majorana fermions (and hence Hilbert space dimension 32), which are simple enough to simulate using existing quantum processors (and very easy to simulate classically). 

Three such models are described in \cite{experiment}, in which the number of terms in the Hamiltonian are 5, 6, and 8 respectively out of the $\binom{10}{4}=210$ possible 4-fermion terms for $N=10$. Therefore, the effective values of the sparsity parameters of these models are $p=5/210 =.0238$, $p=6/210=.0286$, and $p=8/210=.0381$. We refer to these as models (A), (B), and (C) respectively. The Hamiltonians of the models are listed in Appendix \ref{appendix:models}. 
Models (A) and (B) were selected by a machine-learning algorithm searching for models that emulate the unsparsified SYK model in a teleportation process corresponding to the transmission of a qubit through a traversable wormhole connecting two black holes in bulk JT gravity. Model (C) was selected to have a teleportation signal even stronger than the one exhibited by the SYK model.

Our previous results for larger $N$ cannot necessarily be reliably extrapolated to $N=10$, but even for the relatively small Hilbert space dimension of $L=32$, we find qualitatively similar behavior for generic sparsified models. For $N=10$, the appropriate RMT ensemble is GUE, for which RMT predicts a gap ratio $r=0.5996$; this value is accurately reproduced by computations for the $N=10$ model with $p=1$. Studies of relative error for the SFF indicate that the $N=10$ sparsified SYK model has spectral properties adhering to RMT predictions as $p$ decreases until the transition point $p_1 \approx 0.3$ is reached. As the model is sparsified further, we find that $r$ deviates from RMT predictions when $p$ is below the transition point $p_2\approx 0.1$. Note that these transitions occur for models that are considerably less sparse than the learned models (A), (B), and (C). 

We also computed $r$ for models (A), (B), and (C); the results are summarized in Table \ref{tab:N=10-learned}. Since these models are learned by minimizing a physically motivated loss function, it would not be surprising if their properties differ from generic models with the same level of sparsity and indeed we find that to be the case. Consider for example the generic models with $p=.0238$, the same sparsity as the learned model (A). We generated 90,000 samples for this value of $p$; the spectral degeneracies and corresponding values of $r$ we found are summarized in Table \ref{tab:N=10-r}. For model (A), we find 4-fold degeneracy and $r=0.5178$, a value below the RMT prediction $r=0.5996$, but well above the value $r=0.34$ exhibited by generic models with this level of sparsity and degeneracy. 

\begin{table}[ht]
    \centering
    \begin{tabular}{|c|c|c|}
    \hline
        Degeneracy & \# samples & $r$ \\
        \hline
         2-fold & 20,072 & 0.30  \\
         4-fold & 38,126 & 0.34  \\
         8-fold & 21,834 & 0.40  \\
         16-fold & 9,356 & N/A  \\
         32-fold & 612 & N/A  \\
         \hline
    \end{tabular}
    \caption{Distribution of degeneracies for the sparse SYK model with $N=10$ and $p=.0238$, based on 90,000 samples, and the $r$ values obtained from averaging over all samples with specified degeneracy.}
    \label{tab:N=10-r}
    
\end{table} 

Model (B) has 2-fold degeneracy and $r=0.5681$, again below the RMT prediction but well above the value $r=0.31$ for generic 2-fold degenerate models with $p=.0286$. On the other hand, Model (C) has 2-fold degeneracy and $r=0.3240$, roughly comparable to the value $r=0.33$ we find for 
generic models with 2-fold degeneracy and $p=.0381$ (though much different from the value $r=0.600$ found for the unsparsified model). 

\begin{table}[ht]
    \centering
    
    \begin{tabular}{|c|c|c|c|c|}
    \hline
        Model & Degeneracy & $r$ & typical $r$ & RMT $r$\\
        \hline
         (A) & 4-fold & 0.518 & 0.34 & 0.600 \\
         (B) & 2-fold & 0.568 & 0.31 & 0.600  \\
         (C) & 2-fold & 0.324 & 0.33 & 0.600\\
         \hline
    \end{tabular}
    
    \caption{Values of the gap ratio $r$ for the learned models (A), (B), (C), compared with values for typical models with the same degeneracy and sparsity and with the RMT prediction.}
    \label{tab:N=10-learned}
    
\end{table} 

We also computed the SFF for the learned models (A), (B), and (C). Even with Gaussian filtering in place, the SFF for a single sample (without any averaging over samples) has large oscillations, making it difficult to draw conclusions about the spectral properties of a particular sample. But one can argue \cite{cotler2017chaos} that time-averaging for a single sample produces an SFF that closely resembles one obtained by averaging over many samples without time-averaging. Therefore we averaged the SFF for each of the models in small nonoverlapping time windows. The window size was chosen to contain sufficiently many data points to average cleanly, but kept small enough to prevent discretization from obscuring the shape of the curve. 

When we choose a single sample for the unsparsified model with $p=1$, the time-averaged SFF does indeed exhibit a clear ramp and plateau, resembling what we find for the ensemble-averaged SFF. But for the highly sparsified model (A) with effective sparsity $p=.0238$, no ramp is seen, as would be expected for a generic model with $p<p_2$. Similarly, the time-averaged SFF has no discernable ramp for models (B) and (C). 

\begin{figure}[ht]
    \centering
    
    \includegraphics{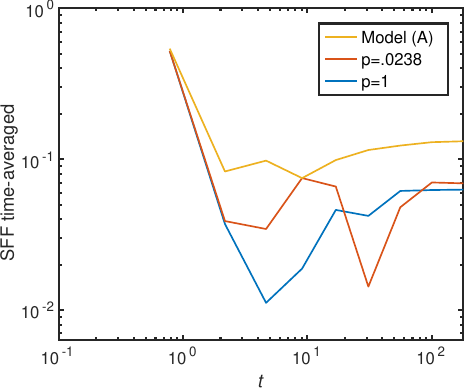}
    
    \caption{Time-averaged SFF (with 10 time windows) for one typical sample of the $N=10$ SYK model with $p=1$, one typical sample with $p=.0238$, and the learned model (A). A ramp is clearly visible for $p=1$, but not for the highly sparsified models.}
    \label{fig:3TWavgsoverlaid}
    
\end{figure}

Interestingly, we find that the learned models are outliers among the ensemble of SYK models with the specified sparsity parameter $p$. In particular, models (A) and (B) have values of the gap ratio $r$ which are significantly larger than the typical value of $r$ in the corresponding ensemble. On the other hand, in all three cases $r$ is well below the RMT prediction, and the time-averaged SFF fails to exhibit a ramp. Thus, since spectral rigidity is compromised at all energy difference scales in the learned models, these models are not good proxies for the unsparsified SYK model, and cannot be expected to capture the dual gravitational behavior of the unsparsified model accurately.

\section{Conclusions}
\label{sec:conclusions}

We have seen that the SYK model can be substantially sparsified while maintaining strong quantum chaos well described by random matrix theory. The spectral rigidity exemplified by RMT is presumed to be necessary for duality with JT gravity to be preserved in the sparsified model. Thus sparsification may open opportunities for instructive classical and quantum simulations of JT gravity requiring substantially reduced computational resources compared to the unsparsified SYK model. 

However, we also found that excessive sparsification generically results in softening of spectral rigidity, signaling that the dual gravitational description breaks down when the SYK Hamiltonian is very sparse. In addition, we studied very highly sparsified nongeneric models selected by a machine learning algorithm \cite{experiment}, and found for those models as well that spectral rigidity is absent. 

A further issue with the learned models is that the dual gravitational description is well established only for the ensemble-averaged SYK model rather than for individual samples. We do not know whether averaging over an ensemble of Hamiltonians is necessary for holography, nor do we know a satisfying bulk gravitational interpretation for the large oscillations in the spectral form factor that occur for individual samples (without time-averaging). Despite such unresolved issues, we expect that strong quantum chaos associated with spectral rigidity is essential for the existence of a useful gravitational dual, and therefore we feel confident that the learned models are not well-suited for describing bulk gravitational phenomena. 

Many questions merit further investigation. A more detailed finite-size analysis might clarify the nature of the transitions at $p=p_1$ (where spectral rigidity begins to soften) and $p=p_2$ (where spectral rigidity is compromised on all energy scales). The effect of sparsification on other quantities, such as fermion correlators and Lyapunov exponents, could be studied. Other values of $q$ (the order of the fermion interactions --- we considered only $q=4$) could be explored, as well as the double-scaled limit in which $N$ and $q$ grow large together \cite{susskind2021entanglement}. Other questions concern the behavior of sparse SYK when $p$ is in between $p_1$ and $p_2$. Is duality with JT gravity maintained for a portion of this regime? Do other instructive gravitational dual descriptions emerge?

Finally, while minimizing a loss function motivated by the behavior of traversable bulk wormholes seems not to unveil a very sparse nongeneric model dual to JT gravity, other training methods might be more successful. For example, one could search for very sparse models such that the spectral form factor, gap ratio, and/or other quantities match well with expectations from random matrix theory. We remain optimistic that suitably sparsified holographic theories can provide powerful experimental access to quantum gravitational phenomena.

\section*{Acknowledgments}
We are grateful for helpful discussions with Brian Swingle, Stephen Shenker, Masanori Hanada, Douglas Stanford, Juan Maldacena, Sandip Trivedi, Alexey Milekhin, Tommy Schuster, and Alexei Kitaev. The research of PO was supported in part by Caltech's Summer Undergraduate Research Fellowship (SURF) program. PO also acknowledges the Perimeter Institute for Theoretical Physics, which sponsored a visit enabling him to present this work at the 2023 It from Qubit Conference. Research at Perimeter Institute is supported by the Government of Canada through the Department of Innovation, Science and Economic Development and by the Province of Ontario through the Ministry of Research, Innovation and Science. HG was supported by the Simons Foundation It from Qubit Collaboration, the Institute for Quantum Information and Matter at Caltech, and BlueQubit Inc. JP acknowledges funding provided by the Institute for Quantum Information and Matter, an NSF Physics Frontiers Center (NSF Grant {PHY}-{1733907}), the Simons Foundation It from Qubit Collaboration, the DOE QuantISED program ({DE}-{SC0018407}), and the Air Force Office of Scientific Research ({FA9550}-{19}-{1}-{0360}). Dedicated in memory of ``Pop'', Lee R.\ Summerlin II (1934--2024), whose everlasting impact as a grandfather, mentor, and teacher reach beyond what words can capture.

\appendix

\section{\boldmath Choice of \texorpdfstring{$\alpha$}{alpha} for Gaussian-filtered spectral form factor}
\label{appendix:alphachoice}

As noted in Sec.~\ref{subsec:filtering}, relevant features of the SFF are more easily seen when a Gaussian filter function is introduced \cite{OnsetOfRMT}. The filter has a tunable parameter $\alpha$ as in \eqref{eq:GaussFiltSFF}, where $\alpha=0$ corresponds to the conventional unfiltered case. Positive $\alpha$ suppresses oscillations in the SFF, improving sensitivity to the time $t_\text{ramp}$ where the ramp begins. Since a shift in the ramp time is a useful diagnostic for non-maximal chaos, filtering is an important element or our analysis of the SFF.

However, applying too much filtering is undesirable, because it distorts the SFF in addition to suppressing its oscillations. In particular, excessive filtering shifts the SFF upward, and pushes the SFFs for different values of $p$ closer together; this reduces the precision of the relative error method we use to estimate $t_\text{ramp}$. Therefore, we carefully choose by hand the lowest $\alpha$ value at which the noisy oscillations are substantially filtered away, thus avoiding a troublesome upward shift. 

\begin{figure}[b!]
    \centering
    
    \begin{subcaptiongroup}
    \centering
    \parbox{0.3\textwidth}{\centering \includegraphics[scale=0.6]{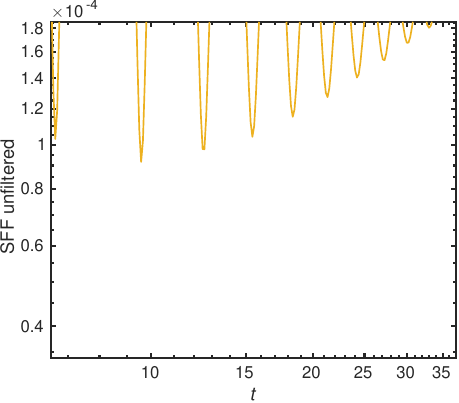} \caption{$\alpha=0.0$}}
    \parbox{0.3\textwidth}{\centering \includegraphics[scale=0.6]{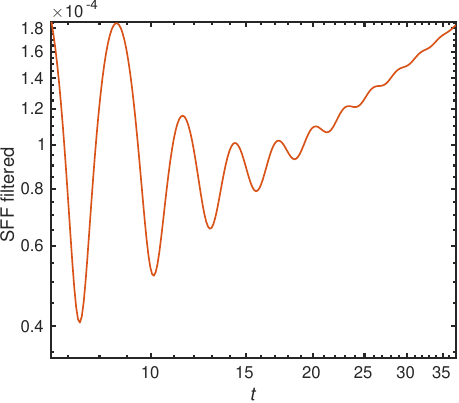} \caption{$\alpha=1.6$}}
    \parbox{0.3\textwidth}{\centering \includegraphics[scale=0.6]{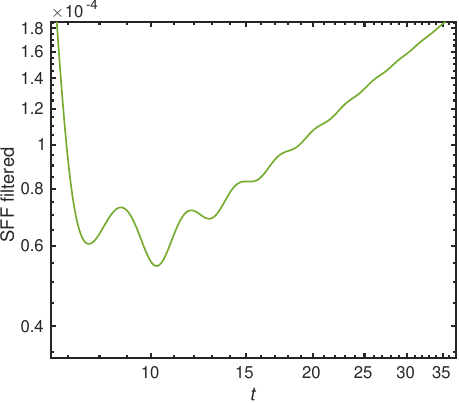} \caption{$\alpha=2.4$}}
    \end{subcaptiongroup}

    \begin{subcaptiongroup}
    \centering
    \parbox{0.3\textwidth}{\centering \includegraphics[scale=0.6]{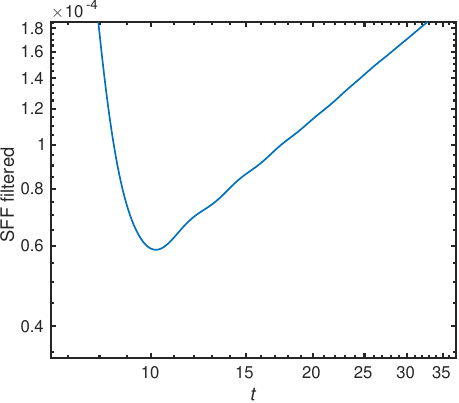} \caption{$\alpha=3.2$}}
    \parbox{0.3\textwidth}{\centering \includegraphics[scale=0.6]{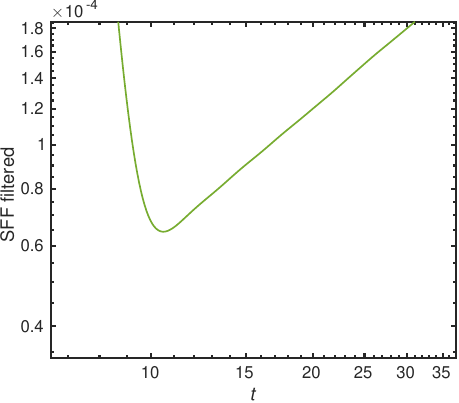} \caption{$\alpha=3.8$}}
    \parbox{0.3\textwidth}{\centering \includegraphics[scale=0.6]{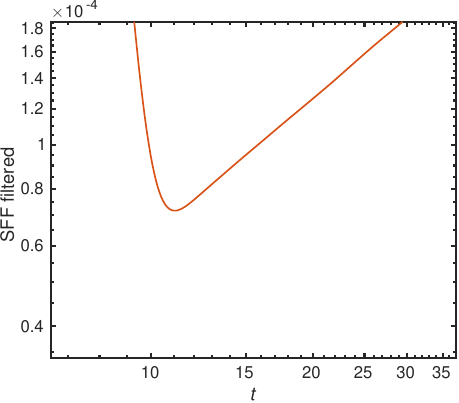} \caption{$\alpha=4.4$}}
    \end{subcaptiongroup}
    
    \caption{SFF vs.~$t$ for the SYK model with $N=18$ and $p=1$, for various values of the Gaussian filtering parameter $\alpha$. Increasing $\alpha$ reduces oscillations, making the onset of the ramp more evident.}
    \label{fig:gaussfilt-illustration}
    
\end{figure}

How the onset of the SFF ramp is more clearly revealed by increasing $\alpha$ is illustrated in Fig.~\ref{fig:gaussfilt-illustration}; in this case our preferred value is $\alpha=3.8$. Gaussian filtering of the SFF was also considered in \cite{OnsetOfRMT}; for each $N$, they chose a value of $\alpha$ similar to ours. 

\section{Sparse SYK models from  machine learning}
\label{appendix:models}

We applied our spectral rigidity tests to three sparse SYK models (denoted here as models (A), (B), and (C)) which were investigated in \cite{experiment}.  These models were obtained by applying a machine learning (ML) algorithm to sparsify a single sample of the $N=10$ SYK model. Models (A) and (B) were selected to emulate the unsparsified SYK model in a teleportation process corresponding to the transmission of a qubit through a traversable wormhole connecting two black holes in bulk JT gravity. Model (C) was selected to have a teleportation signal even stronger than the one exhibited by the SYK model. Properties of model (A) were discussed further in \cite{lykken2024long}.

The Hamiltonian of model (A) contains only 5 of the 210 terms in the full SYK Hamiltonian and involves only 7 out of the 10 Majoranas. These five terms are mutually commuting. The teleportation process in this model was simulated on a quantum processor, as had been proposed in \cite{QGITL1,QGITL2, gao2021traversable}; see also \cite{quantinuumexperiment}. 

\begin{align}
\begin{split}
\textbf{A: } 
H = &-0.36 \psi^1 \psi^2 \psi^4 \psi^5 \\
    &+0.19 \psi^1 \psi^3 \psi^4 \psi^7 \\
    &-0.71 \psi^1 \psi^3 \psi^5 \psi^6 \\
    &+0.22 \psi^2 \psi^3 \psi^4 \psi^6 \\
    &+0.49 \psi^2 \psi^3 \psi^5 \psi^7 .
\end{split}\label{HamA}
\end{align}

In the supplementary material of \cite{experiment} two other models (B) and (C) are reported. (B) is learned the same way as in the (A), but has 6 terms and involves 8 of the Majoranas. Model (C) has 8 terms and involves all 10 Majoranas. Models (B) and (C), unlike model (A), have some noncommuting terms. 

\begin{align}
\begin{split}
\textbf{B: } 
H = &-0.35 \psi^1 \psi^2 \psi^3 \psi^6 \\
    &+0.11 \psi^1 \psi^2 \psi^3 \psi^8 \\
    &-0.17 \psi^1 \psi^2 \psi^4 \psi^7 \\
    &-0.67 \psi^1 \psi^3 \psi^5 \psi^7 \\
    &+0.38 \psi^2 \psi^3 \psi^6 \psi^7 \\
    &-0.05 \psi^2 \psi^5 \psi^6 \psi^7 .
\end{split}
\end{align}

\begin{align}
\begin{split}
\textbf{C: } 
H = &+0.60 \psi^1 \psi^3 \psi^4 \psi^5 \\
    &+0.72 \psi^1 \psi^3 \psi^5 \psi^6 \\
    &+0.49 \psi^1 \psi^5 \psi^6 \psi^9 \\
    &+0.49 \psi^1 \psi^5 \psi^7 \psi^8 \\
    &+0.64 \psi^2 \psi^4 \psi^8 \psi^{10} \\
    &-0.75 \psi^2 \psi^5 \psi^7 \psi^8 \\
    &+0.58 \psi^2 \psi^5 \psi^7 \psi^{10} \\
    &-0.53 \psi^2 \psi^7 \psi^8 \psi^{10} .
\end{split}
\end{align}

\bibliography{cites}
\bibliographystyle{unsrt}

\end{document}